\documentclass[10pt,sigconf]{acmart}
\pdfoutput=1
\AtBeginDocument{%
  }

\setcopyright{acmlicensed}
\copyrightyear{2018}
\acmYear{2018}
\acmDOI{XXXXXXX.XXXXXXX}

\acmConference[Conference acronym 'XX]{Make sure to enter the correct
  conference title from your rights confirmation emai}{June 03--05,
  2018}{Woodstock, NY}

\acmISBN{978-1-4503-XXXX-X/18/06}

\acmSubmissionID{51}

\usepackage{algorithmic}
\usepackage{multirow}
\usepackage{makecell}
\usepackage{tabularx}
\usepackage{float}
\usepackage{color}
\usepackage{algorithm}
\begin{document}

\title{Scene Understanding Enabled Semantic Communication  with Open Channel Coding}
\author{Zhe Xiang, Fei Yu, Quan Deng, Yuandi Li, Zhiguo Wan}

% \renewcommand{\shortauthors}{Trovato et al.}

%%
%% The abstract is a short summary of the work to be presented in the
%% article.
\begin{abstract}
As communication systems transition from symbol transmission to conveying meaningful information, sixth-generation (6G) networks emphasize semantic communication. This approach prioritizes high-level semantic information, improving robustness and reducing redundancy across modalities like text, speech, and images. However, traditional semantic communication faces limitations, including static coding strategies, poor generalization, and reliance on task-specific knowledge bases that hinder adaptability.

To overcome these challenges, we propose a novel system combining scene understanding, Large Language Models (LLMs), and open channel coding, named \textbf{OpenSC}. Traditional systems rely on fixed domain-specific knowledge bases, limiting their ability to generalize. Our open channel coding approach leverages shared, publicly available knowledge, enabling flexible, adaptive encoding. This dynamic system reduces reliance on static task-specific data, enhancing adaptability across diverse tasks and environments. Additionally, we use scene graphs for structured semantic encoding, capturing object relationships and context to improve tasks like Visual Question Answering (VQA). Our approach selectively encodes key semantic elements, minimizing redundancy and improving transmission efficiency. Experimental results show significant improvements in both semantic understanding and efficiency, advancing the potential of adaptive, generalizable semantic communication in 6G networks.
\end{abstract}

%% A "teaser" image appears between the author 

% \received{20 February 2007}
% \received[revised]{12 March 2009}
% \received[accepted]{5 June 2009}

%%
%% This command processes the author and affiliation and title
%% information and builds the first part of the formatted document.
\maketitle

\section{Introduction}
Communication systems have evolved from transmitting mere symbols toward conveying meaningful information to meet the growing demand for high-speed, low-latency, and reliable data transmission. As the development of sixth-generation (6G) networks advances, the focus has shifted from ensuring bit-level accuracy, as per Shannon’s classical theory~\cite{shannon1948mathematical}, to semantic communication ~~\cite{yang2022semantic},~\cite{zhang2022toward},~\cite{9530497}. This paradigm aims to capture, transmit, and reconstruct high-level semantic information relevant to the intended communication tasks. Semantic communication systems are gaining prominence in various modalities, including text ~\cite{xie2021deep}, speech ~~\cite{9500590}, images ~\cite{9685667}, and video ~\cite{9953110}, as they significantly reduce data redundancy and improve transmission robustness by prioritizing meaning over exact bit-level replication.

\begin{figure}
    \centering
    \includegraphics[width=\linewidth]{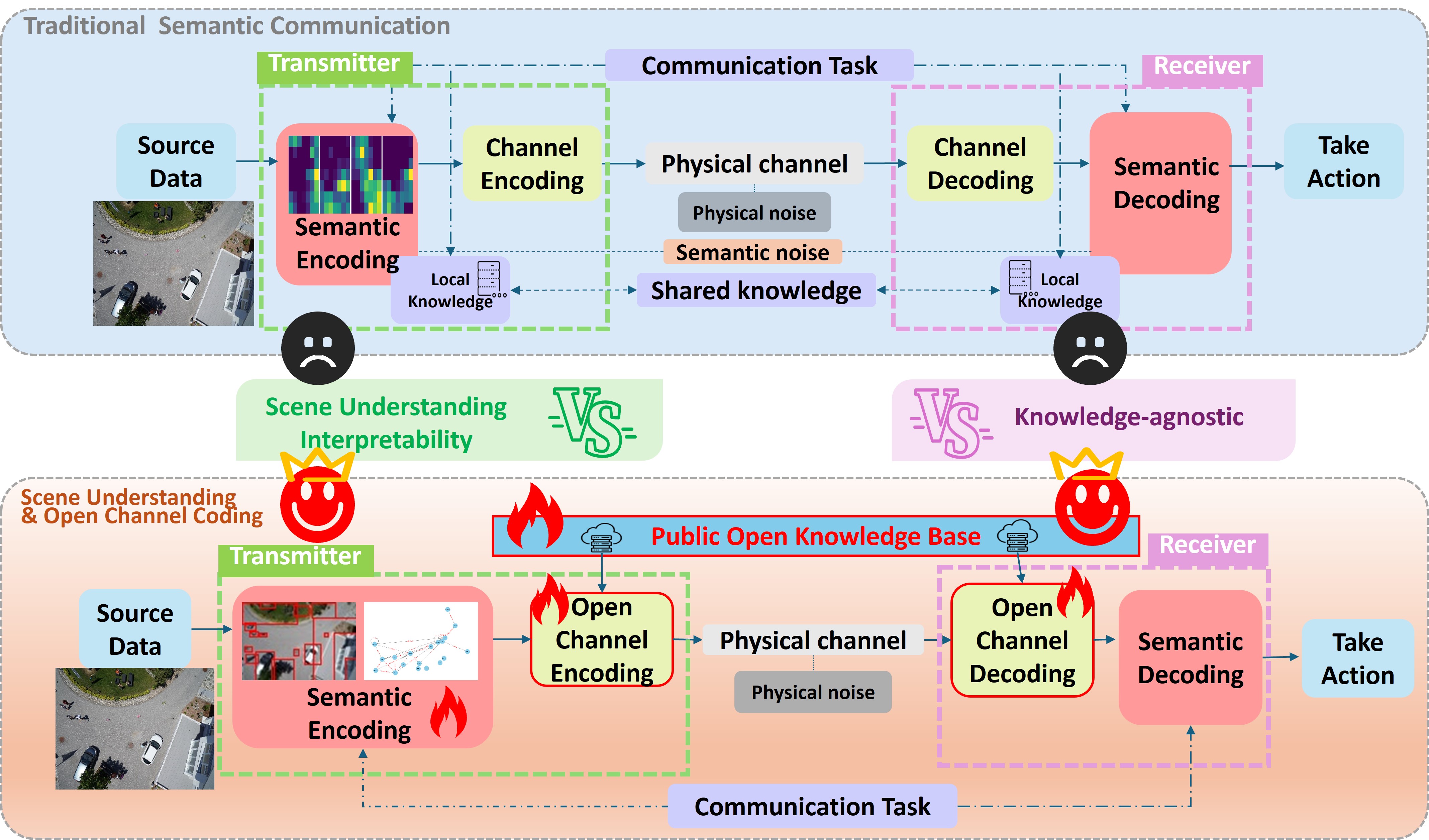}
    \caption{The Comparison of a semantic communication framework with visual scene understanding, incorporating semantic coding and knowledge-agnostic channel coding, with traditional semantic communication approaches ~\cite{yang2022semantic}.}
    \label{fig:intro-figure}
\end{figure}

Large Language Models (LLMs) ~\cite{devlin-etal-2019-bert}, such as ChatGPT ~\cite{Radford2018ImprovingLU}, have demonstrated substantial progress in natural language understanding, offering unprecedented potential for semantic encoding and decoding. By leveraging these models, semantic communication systems can better align transmitted content with the intended meaning, paving the way for more intelligent and adaptive communication across various scenarios. Moreover, visual information structured into scene graphs ~\cite{tang2020unbiased} provides interpretable semantic representations, capturing objects and their relationships. Integrating these structured semantics into communication systems enables improved discrimination and more effective task-oriented performance, such as image retrieval ~\cite{10.1145/3699715} and decision-making tasks ~\cite{raparthi2021real}.

\textbf{Current Issues.} Despite these advancements, existing semantic communication systems face critical limitations: 1) Lack of Visual Semantic Understanding~\cite{zhang2021survey}: Traditional systems primarily focus on encoding low-level image features, neglecting the high-level visual semantic information such as object relationships and contextual meaning. This oversight limits the performance of tasks like Visual Question Answering (VQA) and image retrieval, where understanding the scene's semantics is crucial. 2) Static Knowledge Bases: Most systems rely on static, task-specific knowledge bases that hinder adaptability and generalization across diverse tasks and dynamic environments~\cite{xie2021deep, luo2022semantic, 9953076, dong2022semantic, 9653664, xie2022task, 9877924}. These fixed structures restrict the system's ability to handle novel or evolving communication scenarios, making them less flexible in the context of 6G networks, which demand adaptive solutions. 3) Inefficient Coding and Redundancy: Existing systems often transmit redundant semantic information, leading to inefficient use of bandwidth and resources. Static encoding strategies, such as fixed-length symbols, fail to adapt to varying channel conditions, exacerbating the issue of redundant transmission and reducing overall communication efficiency~\cite{6773395}.

% %%%%%%%%%%%%%%%
% 1) Limited Interpretability ~\cite{zhang2021survey}: Traditional systems rely on black-box deep learning models, leading to poorly interpretable semantic features and reducing system control and flexibility. 2) Static Coding Strategies: Most systems employ static neural networks that generate fixed-length semantic symbols ~\cite{xie2021deep,luo2022semantic,9953076,dong2022semantic,9653664,xie2022task,9877924}, which are unable to adapt to varying channel conditions, resulting in inefficient bandwidth use ~\cite{6773395}. 3) Inefficient Feature Extraction: Redundant semantic information may be transmitted, wasting transmission resources and reducing system efficiency. The need for both interpretable semantic representations and adaptive communication strategies motivates the integration of LLM-based encoding and scene graph-based semantic structures into semantic communication systems.

\textbf{Challenges.} Addressing the limitations of existing semantic communication systems presents the following key challenges:
\begin{itemize}
\item \textit{How to integrate high-level visual semantics} (e.g., object relationships and context) into encoding for tasks like Visual Question Answering (VQA) and image retrieval?
\item \textit{How to develop adaptive encoding mechanisms} that efficiently compress semantic information and optimize transmission under varying channel conditions?
\item \textit{How to enable systems to adapt to dynamic, task-specific scenarios} by leveraging open, shared knowledge instead of static, task-specific knowledge bases?
\item \textit{How to use structured semantic representations} (e.g., scene graphs) to improve interpretability and enhance performance in task-specific applications?
\end{itemize}

To clearly outline the existing issues in current semantic communication systems, Figure~\ref{fig:intro-figure} illustrates the limitations of traditional semantic communication frameworks. To address these issues, we propose a solution that involves design improvements in both the semantic encoding and channel coding components. The main goal is to enhance the interpretability of visual scene understanding within the semantic encoding phase, particularly for visual data, and to overcome the limitations found in existing research regarding the construction of shared knowledge bases in the channel coding phase. By focusing on these aspects, we aim to improve the flexibility and adaptability of semantic communication systems.

\textbf{Proposed Research Solutions.}
To address the challenges in developing a semantic communication system with LLMs and structured semantic encoding, we propose a comprehensive strategy with the following key components: (i) an innovative multimodal semantic communication system that integrates Large Language Models (LLMs) with structured scene graph encoding, enhancing the representation of complex visual-textual relationships for tasks such as Visual Question Answering (VQA). (ii) To optimize transmission efficiency, we introduce a dynamic open channel coding mechanism that adapts in real-time to channel conditions, overcoming the limitations of static knowledge-based systems. (iii) Our selective scene graph encoding prioritizes critical objects and relationships, reducing redundancy and improving the interpretability of semantic representations, enabling more accurate, context-aware responses. (iv) Finally, through comprehensive experiments, we validate our approach, demonstrating significant improvements in semantic-level understanding and transmission efficiency. This strategy enhances the overall efficiency, accuracy, and adaptability of semantic communication systems, particularly in complex multimodal applications like VQA~\cite{qian2022scene,9965282,10073911,9965298}, and supports the shift from centralized to decentralized knowledge bases, enabling local modeling of meaning~\cite{yang2022semantic} while incorporating privacy-aware mechanisms.
% To address the issues and challenges of developing a semantic communication system with LLMs and structured semantic encoding, a comprehensive strategy can be implemented. This strategy emphasizes the integration of LLM-driven semantic encoding that adapts to diverse tasks and data sources, effectively minimizing the reliance on a static shared knowledge base while enhancing the alignment between semantic and technical processes. By employing real-time adaptive channel coding, the system can continuously monitor channel conditions and dynamically adjust its encoding and transmission strategies, optimizing efficiency in varying environments. Additionally, introducing selective scene graph encoding allows for the prioritization of relevant objects and relationships specific to each task, thereby reducing redundancy while ensuring the essential semantic features are preserved. The use of structured semantic representations through scene graphs enhances interpretability and performance in complex multimodal applications like Visual Question Answering (VQA) ~\cite{qian2022scene}. Furthermore, transitioning from a centralized shared knowledge base to a decentralized approach enables local modeling of meaning ~\cite{yang2022semantic}, incorporating privacy-aware mechanisms that secure sensitive information during transmission. This multifaceted strategy aims to enhance the overall efficiency, accuracy, and adaptability of semantic communication systems, particularly when dealing with complex multimodal data.

\textbf{Summary of Novel Contributions.} Our contributions are summarized as follows:
% \linespread{0.5}
\begin{itemize}
\item Multimodal Semantic Integration: We use Large Language Models (LLMs) with structured scene graph encoding. This improves the representation of visual-textual relationships for tasks like Visual Question Answering (VQA).
\item Dynamic Channel Coding: We introduce dynamic open channel coding. It adapts to real-time channel conditions and improves transmission efficiency, overcoming static knowledge-base limitations.
\item Selective Scene Graph Encoding: Our selective encoding prioritizes relevant objects and relationships. This reduces redundancy and improves interpretability for more accurate, context-aware responses.
\item Experimental Validation: Through experiments, the performances show significant improvements in both semantic understanding and transmission efficiency, proving the practical potential of our approach for real-world applications.
\end{itemize}
\section{Related Work}
\subsection{Semantic Communications}
Since the introduction of semantic communication, it has attracted significant attention from both industry and academia, and has been identified as one of the core challenges in wireless communications~\cite{tong2022nine}. Semantic communication aims to accurately transmit semantic information between the sender and receiver~\cite{bao2011towards}. By leveraging advanced AI techniques, it can extract and convey the most relevant information, thereby enhancing transmission efficiency, reducing redundancy, and minimizing delays~\cite{strinati20216g},~\cite{luo2022semantic}. Owing to the development of deep learning in semantic communication, it exhibits superior performance compared to traditional communication methods.

For natural language processing, H. Xie et al.~\cite{xie2021deep} proposed deepSC for text transmission, which aims to maximize the reduction of semantic errors by restoring the semantics of sentences. For image processing, Huang et al.~\cite{9953076} proposed an image semantic communication system that uses Generative Adversarial Networks (GAN) to extract global image semantics and reconstruct images at the receiver. Yufei Bo et al.~\cite{bo2024joint} introduced a joint coding modulation (JCM) framework for digital semantic communication using Variational Autoencoders (VAE). The transformation probabilities from source data to discrete constellation symbols were learned. The semantic communication system proposed by Chen et al.~\cite{dong2022semantic} is characterized by achieving ``intelligent flow" through model propagation, introducing the concept of the Semantic Slicing Model (SeSM).Li et al. ~\cite{li2024multimodal}proposed a cross-modal enhancement method for semantic communication as well as a trustworthy semantic communication framework.

\begin{figure*}[!t]
    \centering
    \includegraphics[width=0.85\linewidth]{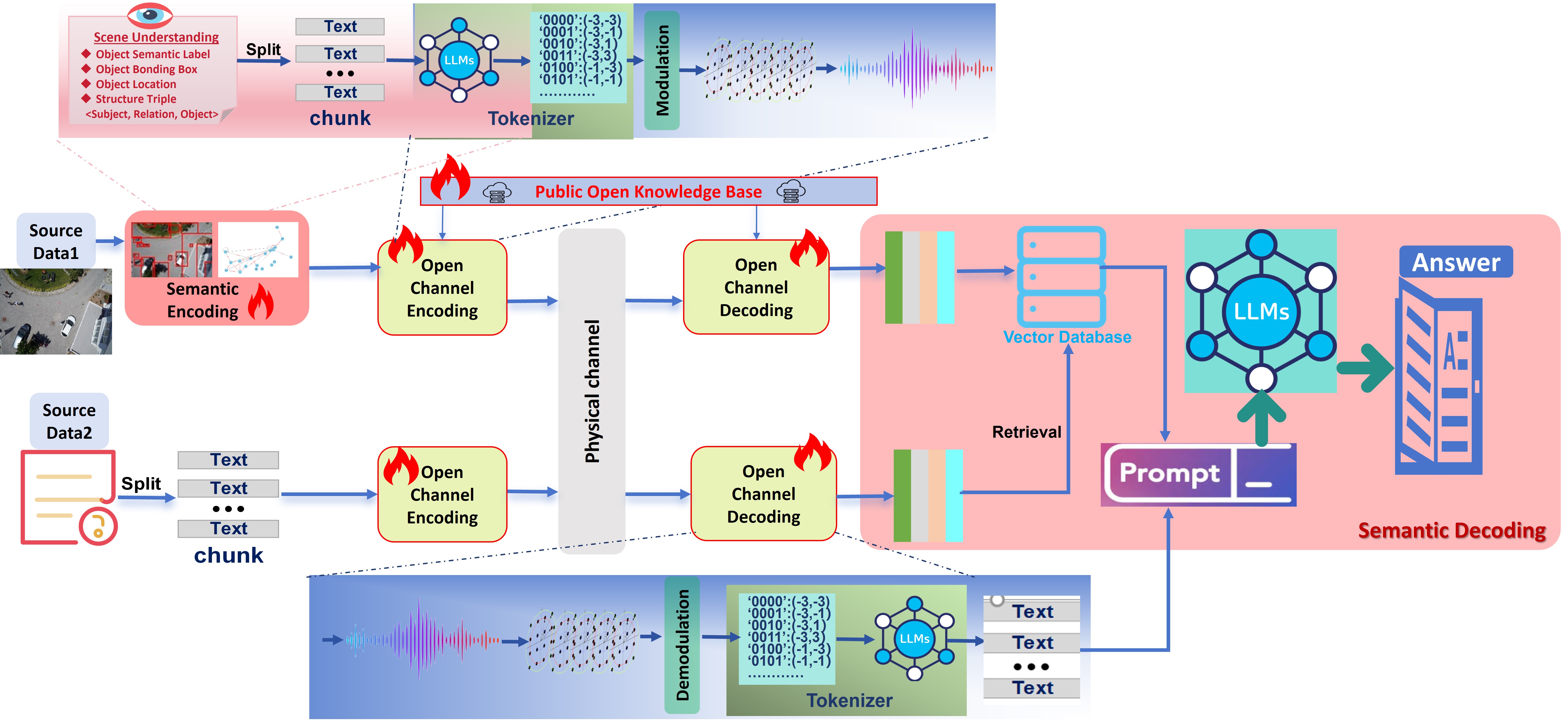}
    \caption{The overall framework of scene understanding-enabled semantic communication with open-channel coding, consisting of three modules: semantic encoding, open-channel coding (encoding and decoding), and semantic decoding. The section marked as "Flame" represents the core contribution of this paper, where we propose a structured semantic encoding method to address challenges in visual information, scene understanding, and interpretability. By leveraging LLMs, we address the problem of knowledge-agnostic open-channel coding.}
    \label{fig:framework}
\end{figure*}

Semantic communication has shown potential in single-modal, single-user scenarios, but it still faces certain limitations in multi-modal contexts. Although previous work has explored some aspects of multi-modal communication, for instance, Xie et al.~\cite{9653664} investigated a multi-user system for visual question answering (VQA) that transmits both images and text. Xie et al.~\cite{xie2022task} introduced a method for text and image transmission based on the Transformer architecture and proposed several deep learning-based frameworks for tasks such as image retrieval and machine translation. Li et al.~\cite{9877924} presented a cross-modal paradigm that leverages complementary information to enhance communication reliability. Luo et al.~\cite{9921202} specifically developed a multi-modal data fusion scheme tailored to the characteristics of wireless channels. However, challenges remain, including the difficulty in constructing a shared knowledge base, the loss and distortion of semantic details during the semantic encoding process, and the issue of generalization across different domains.

Recent advancements in LLMs have significantly impacted semantic communication across various fields, offering potential solutions to the aforementioned issues. For example, Jiang et al.~\cite{10670195} addressed the challenges in semantic communication of image data by incorporating a framework that includes a knowledge base based on the Segment-Anything model, attention-based semantic integration, and adaptive compression techniques. Similarly, proposed a semantic communication framework (LAM-SC) tailored for image data, utilizing LLMs as the core knowledge base~\cite{jiang2024large}. These approaches leverage the profound understanding of human knowledge by LLMs to build robust knowledge bases for different communication tasks. Shen et al.~\cite{shen2024large} harnessed the capabilities of LLM in language understanding, planning, and code generation, integrating them with joint learning strategies oriented to tasks and communication-edge. They proposed an efficient and versatile framework for coordinating edge AI models to perform edge intelligence tasks.

However, the application of LLMs in channel coding and decoding for multi-modal semantic communication, as well as in the construction of multi-modal shared knowledge bases, remains underexplored.

% However, the application of LLMs in channel coding and decoding for multi-modal semantic communication, as well as in the construction of multi-modal shared knowledge bases, remains underexplored.
% 阐述semantic communication的发展现状
% 逐渐引出当前做视觉相关的多模态传输存在的问题（语义编解码存在的问题，通过引出现有文献，指出现状，存在问题；再引出的LLM的相关工作（结合“Large Language Model Enabled Semantic Communication Systems”文章的相关工作），目前存在什么研究，主要应对什么任务，主要思路是什么），最后，综合分析我们和这些方法不同

% \linespread{0.4}
\subsection{Visual Question Answering}
The Visual Question Answering (VQA) task involves matching a natural language question posed by a user with a given image, generating an accurate answer by understanding the image content and parsing the question. VQA tasks require learning how to understand image content and answer related questions from a large dataset of image-question-answer pairs~\cite{antol2015vqa}. In our work, the VQA task differs from traditional VQA tasks in that it does not require image-question-answer pairs. Instead, it allows users to pose questions from four different perspectives, with no restrictions on the length or style of the questions, making it a more complex VQA task. This requires a deep understanding of both the image and the textual context, adding to the challenge of the VQA task.

Traditional VQA systems use object detection, image features, and text descriptions. These systems typically extract low-level visual features as semantic content~\cite{yang2016stacked}. However, the semantic representation capabilities of these methods are limited, and they often fail to capture fine details, leading to errors. In contrast, scene graphs provide a more structured and comprehensive representation of scene semantics, enabling better answers to user questions and improved semantic discrimination. Therefore, in our work, we employ scene graph-based semantic encoding for VQA tasks.

% 引出现有研究，参考“Task-Oriented Scene Graph-Based Semantic Communications with Adaptive Channel Coding”中关于“Image Retrieval”的书写方式
\section{System Model}
The scene understanding enabled semantic communication system utilizes structured semantic encoding and open channel coding to facilitate robust transmission and interpretation of multimodal data. In this section, we introduce the overall framework of the multimodal semantic communications with the structured scene semantic coding and large language model-based channel coding, as shown in Figure~\ref{fig:framework}.

% The framework comprises a semantic transmitter and a semantic receiver, each serving distinct yet complementary roles in processing the input data and executing downstream intelligent tasks.

\subsection{Semantic Transmitter}
The semantic transmitter is divided into two primary components: structured semantic encoding and open channel encoding.

\subsubsection{Structured Semantic Encoding}
The structured semantic encoder extracts and encapsulates semantic features from the input source images. This process can be mathematically represented as:
\begin{equation}
a = S(I; \zeta),
\end{equation}
where \( I \in \mathbb{R}^{H \times W} \) denotes the input image with height \( H \) and width \( W \), \( a \in \mathbb{R}^{L_S} \) represents the extracted structured semantic features with length \( L_S \), and \( \zeta \) signifies the trainable parameters of the network. This encoder effectively captures the latent structured semantic information from the images.

\subsubsection{Open Channel Encoding}

Due to constraints such as limited channel capacity and the presence of noise, the structured semantic representation is further processed by the channel encoder to produce transmitted symbols, expressed as:
\begin{equation}
X = C(a; \gamma),
\end{equation}
where \( X \in \mathbb{C}^{L_C} \) denotes the transmitted complex symbols with length \( L_C \), and \( \gamma \) represents the trainable parameters of the channel encoder. This encoder compresses the semantic information, enhancing the robustness of the communication system against channel variations. The resulting symbols \( X \) undergo normalization of power before physical transmission through the wireless channel.

\subsection{Semantic Receiver}
The semantic receiver consists of two key modules: open channel decoding and structured semantic decoding.

\subsubsection{Open Channel Decoding}
The wireless channel receives \( X \) and outputs \( Y \in \mathbb{C}^{L_C} \) as the received symbols, following the transmission model $Y = HX + N$, where \( N \sim \mathcal{N}(0, \sigma^2_n) \) denotes the independent and identically distributed (i.i.d.) Gaussian noise vector, and \( H \in \mathbb{C}^{L_C} \) represents the channel coefficients. In a Rayleigh fading environment, the channel coefficients are modeled as \( H \sim \mathcal{N}(0, 1) \).

To recover the transmitted symbols, the linear minimum mean-squared error (L-MMSE) estimator is applied, yielding the estimated signals \( \hat{X} \in \mathbb{C}^{L_C} \). The estimated symbols are then processed by the channel decoder, which aims to decompress the structured semantic representations while mitigating the effects of channel fading and noise interference. The reconstruction of the structured semantic features is given by:
\begin{equation}
\hat{a} = C^{-1}(\hat{X}; \theta),
\end{equation}
where \( \hat{a} \in \mathbb{R}^{L_S} \) denotes the recovered structured semantic representation and \( \theta \) are the trainable parameters of the channel decoder.

\subsubsection{Scene understanding-oriented LLM-Semantic Decoding}
The structured semantic decoder integrates the recovered structured semantic information to execute the downstream intelligent task, mathematically represented as:
\begin{equation}
q = S^{-1}(\hat{a}; \delta),
\end{equation}
where \( q \) represents the task result and \( \delta \) denotes the trainable parameters of the semantic decoder. In this work, we focus on the downstream task of visual question answering (VQA) to evaluate the performance and efficacy of the proposed multimodal semantic communication framework.
In a scene understanding-oriented framework, the structured semantic decoder plays a pivotal role by interpreting the recovered structured semantic information and applying it to downstream intelligent tasks. For Visual Question Answering (VQA), the structured semantic information encompasses not only the visual elements of the scene (such as objects, relationships, etc.) but also their semantic connections with the natural language question. This process is mathematically represented as:
\begin{equation}
    q = S^{-1}(\hat{a}; \delta),
\end{equation}
where \( q \) represents the result of the downstream task, specifically the answer to a visual question in VQA. \( \hat{a} \) is the input to the decoder, consisting of structured features derived from the multimodal data (e.g., visual and textual information). \( \delta \) denotes the trainable parameters of the semantic decoder, optimized to align with the task requirements for accurate decoding.

In this framework, scene understanding provides structured information about the objects, relationships, and spatial positions in the image, which are crucial for VQA tasks. By linking visual features from the image to the semantics of the natural language question, the structured semantic decoder extracts fine-grained information and generates contextually relevant answers.

\section{Scene Understanding Enabled Semantic Communication System with LLMs Assistance}

In this section, we design a scene understanding-enabled multimodal semantic communication framework with open channel coding, referred to as \textbf{OpenSC}, to perform the VQA task for Scene Understanding. Figure ~\ref{fig:framework} shows the overall framework of our proposed OpenSC.

% \subsection{Semantic Transmitter}
\subsection{Structure Semantic Encoding}

Scene understanding is crucial for Visual Question Answering (VQA), as it provides a structured representation of the visual environment. We propose Structured Semantic Encoding, which converts raw visual data into scene graph-based representations, capturing both objects and their relationships as <subject-predicate-object> triples. This is achieved through a Prototype-based Embedding Network~\cite{zheng2023prototype}, using Faster R-CNN to extract bounding boxes and feature maps. The union of bounding boxes represents the relationships between entities, forming the foundation for semantic decoding in VQA tasks.

The structured semantic coding method generates compact and distinctive representations for subjects, objects, and predicates by combining class-specific prototypes with instance-specific transformations. For each subject \(s\), object \(o\), and predicate \(p\), linear transformations map them into a shared semantic space, where both class prototypes (\(t_s\), \(t_o\), \(t_p\)) and instance-specific adjustments (\(v_s\), \(v_o\), \(u_p\)) are learned. This design captures both the unique attributes of each entity and the shared characteristics within each category. The representations are mathematically expressed as:
\begin{equation}
\begin{aligned}
s &= W_s t_s + v_s, \\
o &= W_o t_o + v_o, \\
p &= W_p t_p + u_p.
\end{aligned}
\label{eq:representations}
\end{equation}
where \(W_s\), \(W_o\), and \(W_p\) are learnable parameters, and \(t_s\), \(t_o\), and \(t_p\) are the class-specific prototypes for the subject, object, and predicate. The instance-specific components \(v_s\), \(v_o\), and \(u_p\) capture the variability within each category.

To align subject-object pairs with corresponding predicates in a shared semantic space, we define the matching function \(F(s, o)\) as:
\begin{equation}
F(s, o) = \text{ReLU}(s + o) - (s - o)^2,
\end{equation}
where \(F(s, o)\) measures the similarity between the subject and object prototypes, ensuring that the relationship between them approximates the predicate \(p\). This function aims to align the entities within the semantic space, facilitating accurate predicate prediction.

To optimize this matching, we introduce Prototype-guided Learning (PL), which utilizes a loss function to minimize the discrepancy between the predicted relationships and their true semantic representations:
\begin{equation}
L_{e\_sim} = -\log \left( \frac{\exp(\langle r, c_t \rangle / \tau)}{\sum_{j=0}^{N} \exp(\langle r, c_j \rangle / \tau)} \right),
\end{equation}
where \(\tau\) is a learnable temperature hyperparameter, and \(r\) represents the relationship to be matched. This loss function encourages the model to align the subject-object pairs with their correct predicates by leveraging the prototype-based representations.

Additionally, to address the challenge of semantic overlap between predicates, we introduce Prototype Regularization (PR). This regularization term promotes the separation of predicates in the semantic space, ensuring that each prototype maintains its distinct identity:
\begin{equation}
L_{r\_sim} = \| S \|_{2,1} = \sum_{i=0}^{N} \left( \sum_{j=0}^{N} s_{ij}^2 \right)^{1/2},
\end{equation}
where $S$ is the cosine similarity matrix between predicate prototypes. By maximizing the distinction between prototypes, PR improves the model's ability to handle ambiguous cases where predicates might otherwise be difficult to distinguish.

\subsection{Open Channel Condings}

In semantic communication systems, addressing the challenges of traditional joint source-channel coding requires a shift towards more efficient and adaptable strategies. In this context, we introduce a novel approach based on structured scene graph encoding for scene understanding, leveraging the strengths of LLMs to tackle issues related to the complexity, generalization, and limitations of knowledge bases in current systems. Our method combines scene understanding with constellation-based token encoding and modulation techniques to optimize both semantic encoding and communication.

The scene understanding module utilizes Scene Graph Generation (SGG) to extract meaningful semantic information from images, represented as tokenized data. This data is first tokenized using the WordPiece algorithm, a subword tokenization method, to break down textual information from scene graphs into smaller, manageable tokens. The WordPiece algorithm, utilizing a vocabulary \(V\) and sentence \(S\), splits each word \(w\) in the sentence into a sequence of subword tokens \(w_1, w_2, \ldots, w_n\), as defined by: $w = w_1 + w_2 + \ldots + w_n$.

The tokenization process maximizes the matching of the substrings with the vocabulary, ensuring efficient segmentation of the sentence. For each word \(w\), we aim to maximize the length of its matched substring from the vocabulary:
\begin{equation}
    \text{Token} = \arg\max_{\substack{w' \subseteq w \\ w' \in V}} |w'|.
\end{equation}

After tokenizing the scene graph information, the token IDs are extracted from a pre-trained BERT model. These token IDs represent specific positions in the predefined vocabulary, which are then converted into \(m\)-bit binary strings. The binary strings are segmented into \(m/n\) substrings, each representing a QAM (Quadrature Amplitude Modulation) symbol:
\begin{equation}
    d = \sum_{i=0}^{m/n-1} b_i \cdot 2^i,
\end{equation}
where \(b_i\) are the bits in the binary string, and \(d\) is the corresponding decimal value for the QAM symbol.

These QAM symbols are mapped onto a constellation diagram for efficient modulation and demodulation in the communication system. The constellation points correspond to the symbols, allowing for robust transmission of semantic information over the communication channel.

% \subsubsection{Modulation and Demodulation Process}

Two modes of operation are defined for receiving the transmitted signals, ensuring the method’s applicability in various scenarios. 
\begin{itemize}
    \item \textit{When Channel State Information (CSI) is available, we use the Zero-Forcing Linear Minimum Mean Square Error (ZF-LMMSE) detector.} This method combines the advantages of Zero-Forcing and Linear Minimum Mean Square Error to eliminate multipath interference while minimizing noise amplification. The received signal \(Y\) is related to the transmitted signal \(X\) by: $Y = H X + N$, 
where \(H\) is the channel matrix, and \(N\) is the noise vector. The ZF-LMMSE detection matrix \(W\) is computed to estimate the transmitted signal:
\begin{equation}
    \hat{X} = W Y = \left( H^H H + \frac{\sigma_n^2}{\sigma_x^2} I \right)^{-1} H^H Y.
\end{equation}

\item  \textit{When CSI is not available, we rely on symbol demodulation by minimizing the Euclidean distance between the received signal point and predefined symbol points on the constellation diagram.} The likelihood of receiving a symbol \(Y\) from the signal \(X\) is given by:
\begin{equation}
\log P(Y \mid X) = -0.5 \cdot \frac{(Y - X)^2}{\sigma_n^2} - 0.5 \cdot \log(2 \pi \sigma_n^2).
\end{equation}
\end{itemize}

After demodulating the symbols, we reconstruct the token IDs and use the pre-trained BERT ~\cite{devlin-etal-2019-bert} model to map these back to the corresponding tokens. The final tokens are combined to reconstruct the original information, ensuring that semantic understanding is preserved during the transmission process. This approach efficiently integrates scene understanding and semantic communication, optimizing both the encoding of visual information and the transmission process, leading to improved performance in tasks like Visual Question Answering (VQA) under variable channel conditions.

\subsection{Scene Understanding-oriented LLM-Semantic Decoding}
In this approach, our main idea is to improve the precision and completeness of the received data, which may be corrupted or incomplete due to transmission issues, by using LLM. 
The key idea is to leverage the semantic knowledge embedded in structured data (such as scene graphs) to correct errors and supplement missing information, ensuring a more reliable and accurate representation of the data in a structured format, such as JSON. 

% When a user query is received, it follows a similar process. The query is reconstructed and corrected using the LLM, then compared with the scene graph to extract relevant information such as object count, location, or relationships. The process is as follows:
% \begin{itemize}
%     \item  Query Reconstruction: The LLM corrects and reconstructs the query from token IDs to a meaningful question.
% \item  Scene Graph Query: Based on the query, relevant structured data is retrieved from the scene graph.
% \item  Preliminary Answer Generation: The LLM uses the retrieved scene graph data to generate an initial response.
% \item  Enhanced Answer: The LLM integrates the context from the query and scene graph to provide a semantically enriched and accurate answer.
% \end{itemize}

\subsubsection{Reconstruction of Initial Information}
The first step involves the conversion of token IDs into readable text. This is done by using a tokenizer to map the numeric token IDs back to their corresponding textual tokens. The output of this step is the reconstructed initial information, which is expressed as:
\begin{equation}
    \mathit{X} = [\textit{In}_1, \ldots, \textit{In}_n],
\end{equation}
where \( \mathit{X} \) represents the sequence of tokens that, when mapped, form the initial information.

\subsubsection{Enhancement with Structured Information}
The reconstructed information may be incomplete or contain errors, so we supplement it with previously structured knowledge. This structured information typically includes object attributes such as quantity, location, and relationships, and is organized in the form of a Scene Graph. This structured data provides additional context, allowing the LLM to correct and enrich the initial information: 
\begin{equation}
\mathit{p} = [\textit{PIn}_1,\ldots, \textit{PIn}_m].
\end{equation}

\subsubsection{LLM-based Semantic Enhancement}
The LLM is responsible for integrating the initial information with the structured data. The model processes both inputs, learning to generate a corrected and enriched version of the data. This process can be described as a series of operations through a multi-layer perceptron (MLP), where the inputs are processed in multiple hidden layers to produce the final output:
\begin{equation}
\mathbf{h}_1 = \sigma(W_1 [\mathit{X}; \mathit{p}] + \mathbf{b}_1),
\end{equation}
where \( W_1 \) and \( \mathbf{b}_1 \) are the weight matrix and bias vector of the first layer, respectively, and \( \sigma \) is the activation function (e.g., ReLU, Sigmoid, or Tanh). Subsequent hidden layers are computed similarly:
\begin{equation}
\mathbf{h}_L = \sigma(W_L \mathbf{h}_{L-1} + \mathbf{b}_L),
\end{equation}
where \( L \) is the number of hidden layers, and \( W_i \) and \( \mathbf{b}_i \) are the weight matrix and bias vector of the \( i \)-th layer, respectively. The final information ($\textit{Final}\_In$) is then obtained from the output layer:
\begin{equation}
\textit{Final}\_In= W_{\text{out}} \mathbf{h}_L + \mathbf{b}_{\text{out}},
\end{equation}
where \( W_{\text{out}} \) and \( \mathbf{b}_{\text{out}} \) are the weight matrix and bias vector of the output layer, respectively. The final information (\(\textit{final\_information} \)) is a vector containing three elements: number, location, and relationship, denoted as \( n \), \( l \), and \( r \):
\begin{equation}
\textit{final\_information} = [n, l, r],
\end{equation}
where \( n \) is the object count, \( l \) is the location, and \( r \) represents the relationships between objects.

\subsubsection{Conversion to JSON Format}
The final enhanced information is converted into a structured JSON format for easy storage and further processing:
\[
\textit{Json\_data} = \left\{
\begin{array}{ll}
\textit{``number"}: n \\
\textit{``location"}: l \\
\textit{``relationship"}: r
\end{array}
\right\}.
\]

This conversion allows for the data to be easily accessed and utilized in subsequent steps.

\subsubsection{Vector Database and Semantic Enhancement Prompt Generation}

Organize objects, relationships, and attributes into chunks representing specific categories with their quantity, location, and relationships. Embed these chunks into a high-dimensional vector space to get dense vector representations. Store these vectors in a temporary database, linking each to its chunk. During retrieval, find the four closest pre-computed vectors with the highest cosine similarities to the query vector. Retrieve the four most relevant chunks and form a prompt to enhance the LLM's query processing. Feed the enhanced prompt into the VQALLM (Visual Question Answering-Large Language Model)\footnote{Utilize existing multimodal large language models to answer our Visual Questions, with the ability to switch out this component as needed to accommodate different requirements, named VQALLM.}.

\subsection{Training Algorithm}
% The training process of the proposed system consists of three modules: visual structure semantic representation, open channel coding, and VQA-oriented LLM-semantic decoding.

% The training process of the proposed system consists of three main modules: visual structured semantic representation, open channel coding, and VQA-oriented LLM-semantic decoding. In this framework, only the structured semantic representation module, which performs scene understanding on visual data (image data) through scene graph generation, requires a pre-trained model. 

% For channel encoding and decoding, we leverage a large language model combined with constellation diagram tokens for symbol mapping, allowing semantic representations to be effectively transmitted over the channel without further training. Therefore, the only module that requires training in our system is the structured semantic encoding part at the transmitter side. The training algorithm for the structured semantic encoding module is described in the pseudocode as shown in 

The training process of the proposed system involves three modules: visual structured semantic representation, open channel coding, and VQA-oriented LLM-semantic decoding. Only the structured semantic representation module, which generates scene graphs from visual data, requires pre-training. Channel encoding and decoding use a large language model with constellation diagram tokens, enabling semantic transmission without additional training. Thus, only the structured semantic encoding at the transmitter requires training, as outlined in Algorithm~\ref{alg:train}.

% \vspace{-0.1cm}
\begin{algorithm}[!h]
\caption{The taining algorithm of Semantic Encodings in our proposed OpenSC.}
\label{alg:train}
\begin{algorithmic}[1]
    \STATE \textbf{Initialization:} The training dataset $\mathcal{K}$.
    \STATE \textbf{Structured Semantic Encoder Training:}
    \STATE \hspace{1em} \textbf{Input:} Sample mini-batch of input images from the training dataset $\mathcal{K}$.
    \STATE \hspace{1em} Extract structured semantic features from images using the encoder: $a = S(I; \zeta)$.
    \STATE \hspace{1em} \textbf{Train} $\zeta$ \textbf{using} gradient descent with the encoder loss.
    \STATE \hspace{1em} \textbf{Compute encoder Loss:} $L = L_{r\_sim} + L_{e\_sim}$.
    
    \STATE \hspace{1em} \textbf{Return:} Trained structured semantic encoder $S(\cdot; \zeta)$.
\end{algorithmic}
\end{algorithm}
% \vspace{0.2cm}
The training time complexity is \( O(N \cdot (H \times W \times C \times L + P)) \), where \( N \) is the dataset size, \( H \), \( W \), and \( C \) are the image dimensions, \( L \) is the encoder depth, and \( P \) is the number of parameters. The complexity is driven by image size, encoder depth, and model parameters.
\section{Experiments}
% In this section, we compare the propsed OpenSC against other methods across against traditional source coding and channel coding methods, as well as mainstream approaches such as DeepSC~\cite{xie2021deep}. Furthermore, the performance of these methods is evaluated across three different digital modulation schemes. Additionally, an ablation study is conducted to carefully examine the contribution of each component within our OpenSC framework. The average number of symbols transmitted by different methods in this task is calculated, and the computational complexity of our method is analyzed.
In this section, we compare the proposed OpenSC with traditional source and channel coding methods, as well as mainstream approaches like DeepSC~\cite{xie2021deep}, in three digital modulation schemes. We also conduct an ablation study to assess the contribution of each component in the OpenSC framework and evaluate the average number of symbols transmitted and the computational complexity of our method.

% \begin{figure*}[!htbp]
%     \centering
% \includegraphics[width=0.85\linewidth]{ACM Sys/@figures/visualization/vis5.jpg}
%     \vspace{-0.5cm}
%     \caption{This is a visualization that clearly demonstrates the differences between our method and other methods.}
%     \label{fig:visualization}
% \end{figure*}

\subsection{Implementation Details}

\subsubsection{Dataset Description}
In the experiments, we use the open source dataset (AUG dataset ~\cite{Li2024AUGAN}), which contains 400 aerial view images of the city. On average, each image includes 63 objects and 42 relationships. Each image has a pixel dimension of 6,000 by 4,000 pixels, which is approximately 24 million pixels (24Mpx). This provides a wealth of detail in each image, making them suitable for high-resolution image analysis and processing, especially in scenarios that require precise identification and analysis of objects within the image. The objects in these images are small and more densely packed, covering 77 object categories and 63 types of relationship. On average, each image includes 63 objects and 42 relationships.

\subsubsection{Evaluation Metrics}
Unlike traditional VQA tasks, which involve a fixed question and corresponding ground truth answer, our approach allows for open-ended questions about the real objects, quantities, locations, and relationships in an image. To evaluate model performance, we use two metrics: recall rate and F1-score. Recall measures the ratio of correctly described real objects, quantities, locations, and relationships in the model’s answers. F1-score combines recall and precision to assess the balance between correct and incorrect object, quantity, location, and relationship descriptions provided by the model.

% In contrast to traditional VQA tasks, which typically involve a question and its corresponding ground truth answer, our approach enables users to ask open-ended questions about the real objects, their quantities, locations, and relationships depicted in an image. To evaluate the performance of each model on the VQA task, we use the following metric: recall, which measures the ratio of correctly described real objects, quantities, locations, and relationships in the image answers returned by a specific model.
% We also use the F1-scores, which combine recall and precision, to further reflect the ratios of incorrect objects, quantities, locations, and relationships in the answers provided by a specific model. 

\subsubsection{Implementation Setting}
The SGG model uses GeneralizedRCNN \cite{he2017mask} with an \texttt{R-101-FPN} backbone \cite{lin2017feature} and ResNet \cite{he2016deep} module, setting the output channel to 256, with 32 groups and a group width of 8. The Region Proposal Network (RPN) \cite{ren2016faster} is configured with FPN, anchors of sizes (32, 64, 128, 256, 512), and strides of (4, 8, 16, 32, 64), with aspect ratios (0.2323, 0.6337, 1.2848, 3.1509). During training, 12,000 candidate regions are selected pre-NMS, and 2,000 post-NMS. Testing uses 6,000 pre-NMS and 1,000 post-NMS regions. The RPN has a middle channel size of 256. The ROI Heads section has a positive sample ratio of 0.5, IoU threshold of 0.3, and a batch size of 256. The ROI Box Head has a pooling resolution of 7, scales of (0.25, 0.125, 0.0625, 0.03125). The number of categories is 77, and the MLP head dimension is 4,096. The training sets a base learning rate to 0.001, weight decay to 0.0001, warmup factor to 0.1, momentum to 0.9, and gradient clipping to 5.0. Learning rate decay occurs sets the 90,000 and 120,000 iterations.

\subsection{Baselines}
For performance comparison of our proposed OpenSC, we consider the following baselines selected from both traditional communication and semantic communication perspectives\footnote{In this comparative experiment, VQALLM  uses the Qwen-plus model for all instances.}.

\noindent \textbf{\textit{Traditional Communication}}: 
% \vspace{-0.2cm}
\begin{itemize}
    \item \textit{Text-based Transmission.} The image’s textual description is converted into pure text. The system uses 5-bit encoding, RS encoding for transmitting text and question information, answering the question at the terminal.
    \item \textit{Text-based Transmission with Huffman Encoding.} The image’s textual description is converted into pure text. Huffman coding, combined with RS encoding, is used to propagate text and question information,  providing answers at the terminal.
    \item \textit{Image Compression with LDPC and Lossless Text Transmission.} The original image is processed using JPEG compression and LDPC encoding. After transmission, the image is reconstructed at the receiver, assuming lossless transmission of the question text, used for answering at the terminal.
\end{itemize}
% \begin{itemize}
% \item 
% The textual description information corresponding to the image is converted into pure text.The combination of 5-bit encoding, RS encoding, and VQALLM for propagating text information and question information describing single-modality images, with VQALLM being used to answer questions at the terminal.
% \item The textual description information corresponding to the image is converted into pure text.Huffman coding combined with RS encoding and VQALLM for propagating text information and question information describing single-modality images, with VQALLM being used to answer questions at the terminal.
% \item Use the original data image without semantic encoding, employ the commonly used JPEG compression algorithm, and process the source data image with 1/2 of LDPC encoding. After transmission through the channel, reconstruct at the receiving end. It is assumed that the user transmits the question text losslessly, and VQALLM is used to answer questions at the terminal.
% \end{itemize}
% \vspace{-0.2cm}
\textbf{\textit{Semantic Communication}}:
% \vspace{-0.2cm}
\begin{itemize}
    \item \textit{Scene Graph Transmission with DeepSC.} Scene graph structure information, extracted via semantic encoding, is converted into plain text. The text and questions are encoded and transmitted using the DeepSC framework, providing answers at the terminal.
    \item \textit{ MJCMSC Framework for Multimodal Transmission.} Using source image and question text data, JCMSC~\cite{10530261} is applied for single-modal image recovery. A new multimodal framework, MJCMSC, is created by adding text transmission capabilities.
    \item  \textit{LLM-SC Framework with Semantic Encoding.} Semantic encoding-extracted scene graph information is converted to plain text, alongside question text data. The LLM-SC~\cite{wang2024large} framework is used for transmission, answering questions at the terminal.
\end{itemize}

\begin{figure*}[!h]
    \centering
    \includegraphics[width=\linewidth]{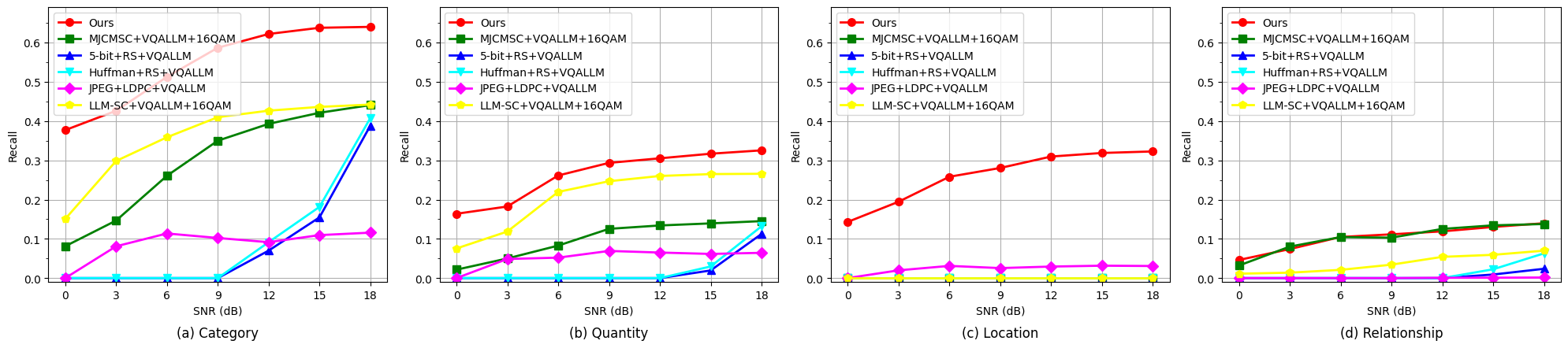}
    \vspace{-0.8cm}
    \caption{Recall performance evaluation under different SNR levels in an AWGN channel using a 16QAM modulation scheme. The MJCMSC system represents our constructed JCMSC multimodal system, where 5-bit + RS Huffman + RS JPEG + LDPC all use 16QAM. The performance is evaluated for four types of questions: Category, Quantity, Location, and Relationship.}

    % \caption{In an AWGN channel under different SNR levels, using a 16QAM modulation scheme (the MJCMSC system is our constructed JCMSC multimodal system, where 5-bit + RS Huffman + RS JPEG + LDPC all use 16QAM), the Recall performance for four types of questions (Category, Quantity, Location, Relationship) is evaluated.}
    \label{fig:16QAM recall}
\end{figure*}
\begin{figure*}[!h]
    \centering
    \includegraphics[width=\linewidth]{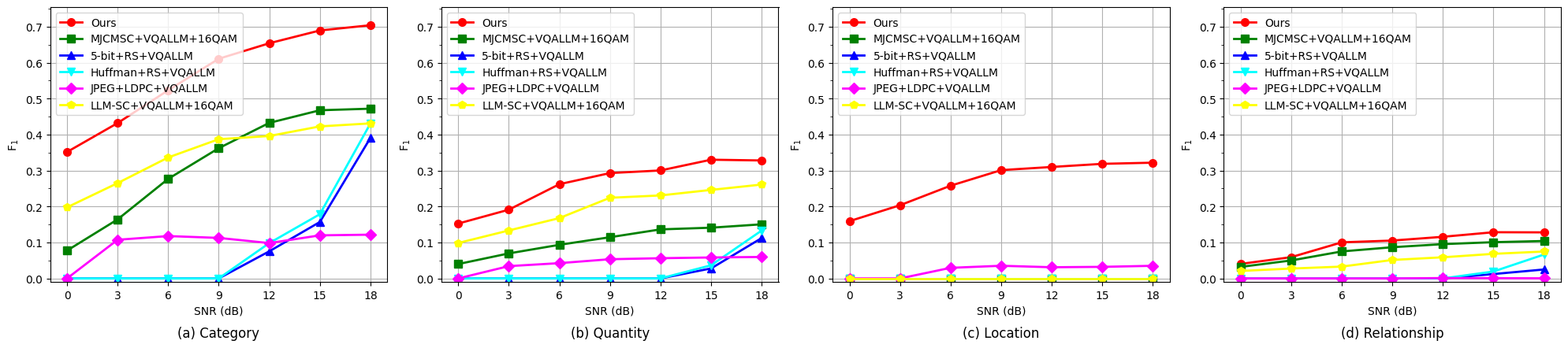}
    \vspace{-0.8cm}
    % \caption{In an AWGN channel under different SNR levels, using a 16QAM modulation scheme, the F1 score for four types of questions (Category, Quantity, Location, Relationship) is evaluated.}
    \caption{F1-score evaluation for four types of questions (Category, Quantity, Location, Relationship) under different SNR levels in an AWGN channel using a 16QAM modulation scheme.}

    \label{fig:16QAM F1}
\end{figure*}

\begin{figure*}[!h]
    \centering
    \includegraphics[width=\linewidth]{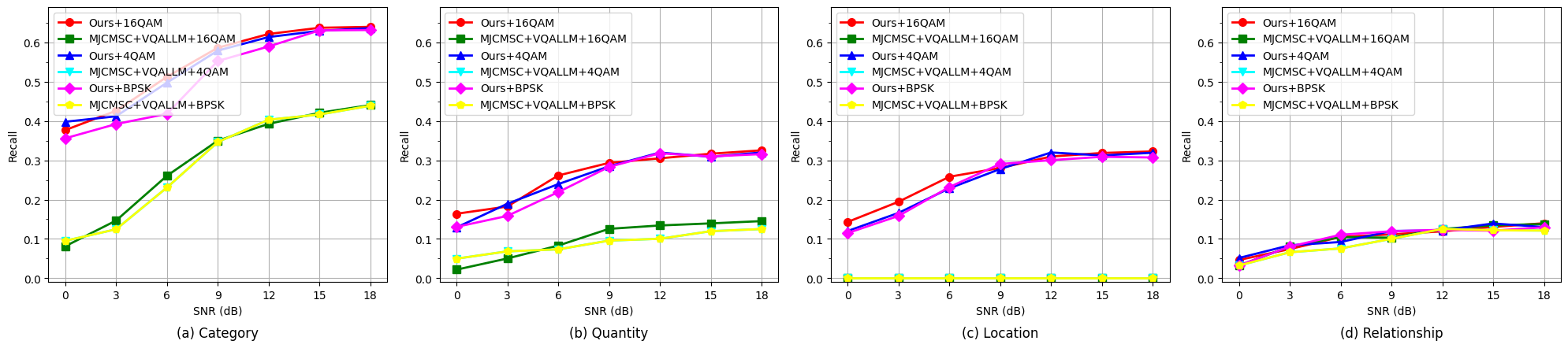}
    \vspace{-0.8cm}
    % \caption{In an AWGN channel under different SNR levels, using three different modulation schemes (BPSK, 4QAM, 16QAM), the recall performance for four types of questions (Category, Quantity, Location, Relationship) is evaluated.}
    \caption{Recall performance evaluation for four types of questions (Category, Quantity, Location, Relationship) under different SNR levels in an AWGN channel, using BPSK, 4QAM, and 16QAM modulation schemes.}

    \label{fig:Comparison recall}
\end{figure*}
\begin{figure*}[!h]
    \centering
    \includegraphics[width=\linewidth]{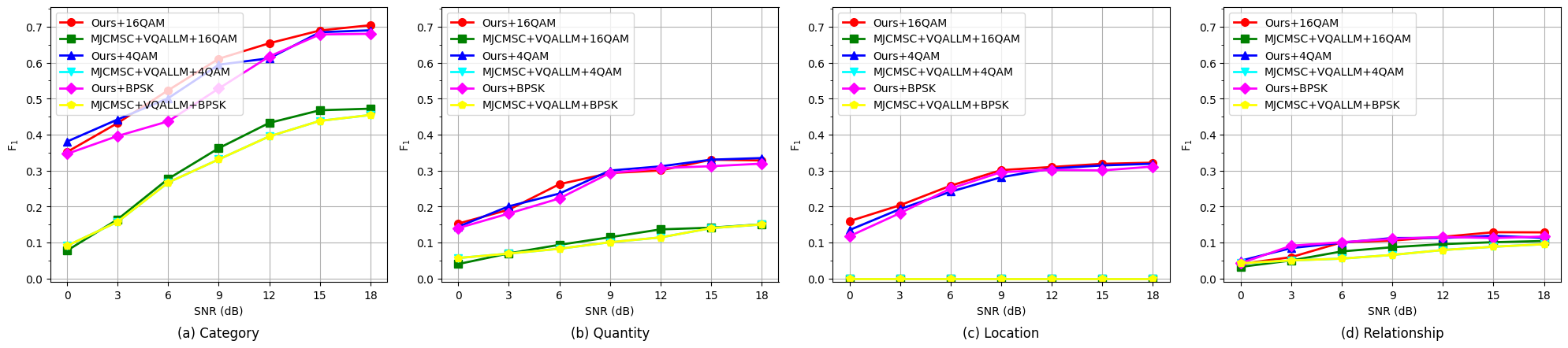}
    \vspace{-0.8cm}
    % \caption{In an AWGN channel under different SNR levels, using three different modulation schemes (BPSK, 4QAM, 16QAM),the F1 score for four types of questions (Category, Quantity, Location, Relationship) is evaluated. }
    \caption{F1-score evaluation for four types of questions (Category, Quantity, Location, Relationship) under different SNR levels in an AWGN channel, using BPSK, 4QAM, and 16QAM modulation schemes.}

    \label{fig:Comparison F1}
\end{figure*}

\begin{figure*}[!h]
    \centering
    \includegraphics[width=\linewidth]{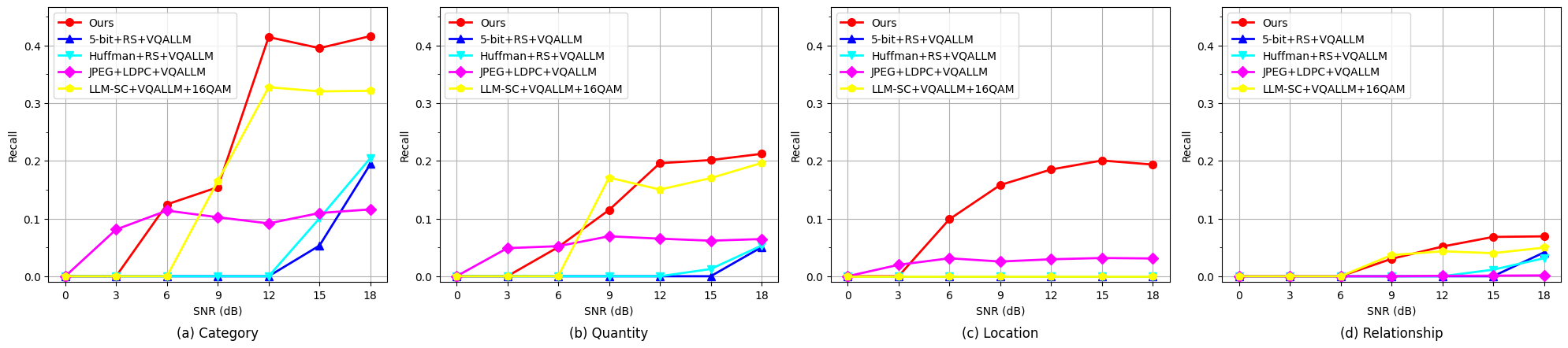}
    \vspace{-0.8cm}
    \caption{Recall performance for four types of questions (Category, Quantity, Location, Relationship) under different SNR levels in a Rayleigh channel using a 16QAM modulation scheme.}

    % \caption{In a Rayleigh channel under different SNR levels, using a 16QAM modulation scheme, the recall performance for four types of questions (Category, Quantity, Location, Relationship) is evaluated.}
    \label{fig:Rayleigh recall}
\end{figure*}
\begin{figure*}[!h]
    \centering
    \includegraphics[width=\linewidth]{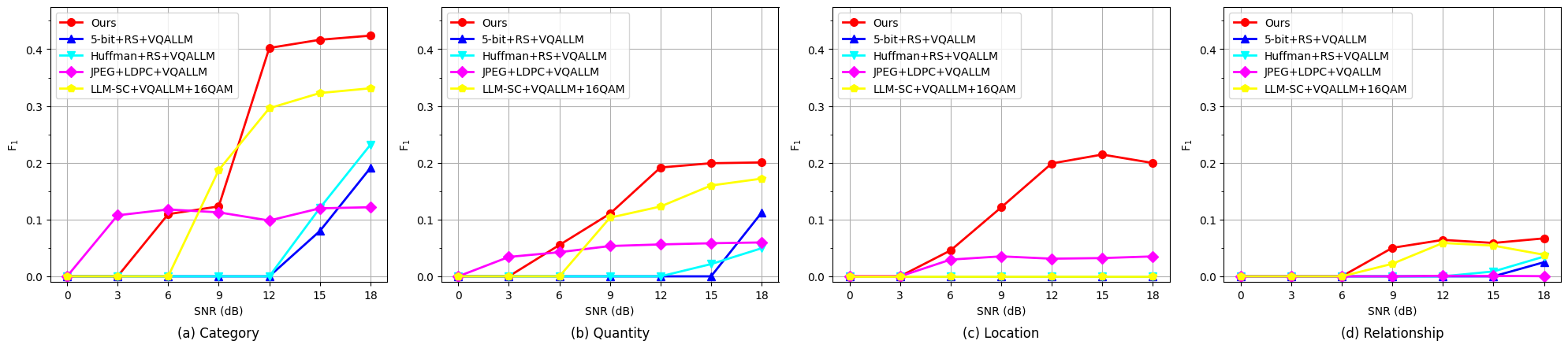}
    \vspace{-0.8cm}
    \caption{F1 score for four types of questions (Category, Quantity, Location, Relationship) under different SNR levels in a Rayleigh channel using a 16QAM modulation scheme.}
    % \caption{In a Rayleigh channel under different SNR levels, using a 16QAM modulation scheme, the F1 score for four types of questions (Category, Quantity, Location, Relationship) is evaluated.}
    \label{fig:Rayleigh F1}
\end{figure*}

\subsection{Experimental Results and Analysis}
\subsubsection{Performance of Different Methods Under Varying Channels}
Figure~\ref{fig:16QAM recall} to Figure~\ref{fig:Rayleigh F1} display the performance of different methods across four VQA question types (Category, Quantity, Location, Relationship). Unlike traditional VQA tasks with one-to-one question-answer pairs, our approach allows for four question types, evaluating performance for each. The experimental results show that DeepSC shows poor performance, with BLUE (1-gram) and Sentence Similarity values under 0.1 across SNRs from 0 to 18 dB (Table~\ref{tab:Deep-SC}). This indicates that DeepSC struggles to recover semantic content when propagating text derived from scene graph structures. Additionally, its poor generalizability limits its application outside of the European Parliament dataset.

\begin{table}[H]
    \centering
    % \renewcommand{\arraystretch}{1.5} % 调整行间距
    % \setlength\tabcolsep{1pt} % 调整列间距
    % \caption{Average BLEU (1-gram) Scores for Different SNR Conditions on the AUG Dataset Using the Deep-SC Method}
    \caption{Average BLEU (1-gram) scores under different SNR conditions on the AUG dataset using the Deep-SC method.}
    \vspace{-0.3cm}
    \label{tab:Deep-SC}
    \resizebox{\linewidth}{!}{
    \begin{tabular}{c|ccccccc}
\toprule
        \textbf{SNR} & \textbf{0dB} & \textbf{3dB} & \textbf{6dB} & \textbf{9dB} & \textbf{12dB} & \textbf{15dB} & \textbf{18dB} \\
        \hline
        \textbf{BLEU (1-gram)} & 0.0190 & 0.0222 & 0.0221 & 0.0240 & 0.0223 & 0.0212 & 0.0215 \\
        % \hline
        \textbf{Sentence Similarity} & 0.0198 & 0.0230 & 0.0235 & 0.0268 & 0.0244 & 0.0231 & 0.0218 \\
\bottomrule
    \end{tabular}
    }
\end{table}

Figure~\ref{fig:16QAM recall}  and Figure~\ref{fig:16QAM F1} compare our method with others under AWGN channel conditions, showing consistent outperformance across SNR levels. Specifically, for location-related questions, methods like MJCMSC, LLM-SC, Huffman, and 5bit score zero, as they lack a mechanism for location description. In contrast, our method calculates object center points and maps them to image regions, embedding structured data into high-dimensional vectors for semantic retrieval and enhancement. Traditional methods like Huffman and 5bit struggle at low SNRs and underperform at high SNRs due to their reliance on pure text, which lacks semantic depth. The JPEG+LDPC method, focused only on image data, fails to enable effective VQALLM-based question answering.

For the Rayleigh channel  Figure~\ref{fig:Rayleigh recall} and Figure~\ref{fig:Rayleigh F1}), our method outperforms others across most SNR levels. While JPEG+LDPC performs well for Category and Quantity questions at low SNR (0-3 dB), and LLM-SC leads around 9 dB, our approach consistently excels for Location and Relationship questions.

\subsubsection{Performance of Different Modulation Methods}
Figure~\ref{fig:Comparison recall} and Figure~\ref{fig:Comparison F1} show the performance comparison of our method in terms of Recall and F1-score under different modulation schemes (BPSK, 4QAM, 16QAM). It can be seen that our performance is quite similar across the three modulation methods, indicating that changing the modulation scheme does not affect the robustness of our system.This indicates that different modulation schemes and encoding lengths can be selected based on varying channel conditions, and the encoding length can be adjusted according to different modulation methods. Furthermore, we have also conducted experiments with the three different modulation methods within the MJCMSC framework, and in all four types of questions, our approach has an advantage.

\subsection{Ablation Experiments}
To validate the effectiveness of the proposed Scene Understanding Enabled Semantic Encoder module, the Open Channel Coding module, and the large language model (LLM) used at the receiver end for the VQA task, we designed three ablation experiments. These experiments focus on evaluating the individual contributions of each module to the overall performance.

\subsubsection{The Ablation Study of Semantic Encoder and Open Channel Coder}
% Additionally, the recall rate for the other three types of questions dropped significantly.After ablating the OCC encoding method, we adopt the Deep-SC encoding and decoding approach. As shown in Table~\ref{tab:Deep-SC}, we unable to recover complete semantic information using this method.When both SGG and OCC are ablated, we attempted a method similar to image recovery, using JCMSC to transmit image information and recover it at the receiver end. The recovered image and the question were then passed to the VQALLM. The results show that this approach performs significantly worse than our method.This is due to the fact that without additional information, VQALLM tends to overlook the deep semantic information in images when only image data is available.
%%

The ablation study underscores the critical importance of both the Structured Semantic Encoder (SSC) and Open Channel Coder (OCC) in achieving high VQA performance. Removing OCC and using the Deep-SC approach significantly degraded the system's ability to recover semantic information, as reflected in the sharp drop in recall rates for ``Category," ``Quantity," and ``Relationship" questions (\textbf{Table}~\ref{tab:Recall adapt}). 
\begin{table}[h]
    \centering
    \caption{Performance comparison across different Large Language Models for each attribute.}
% \caption{Indicating Performance Across Different Large Language Models for Each Attribute}
\vspace{-0.3cm}
    \label{tab:Recall adapt}
    \resizebox{\linewidth}{!}{ % Resizes the table to fit the text width
    \begin{tabular}{ccccccc}
        \toprule
        \multirow{2}{*}{\textbf{VQA LLM}} & \multirow{2}{*}{\textbf{SSC}} & \multirow{2}{*}{\textbf{OCC}} & \multicolumn{4}{c}{\textbf{Performance Metrics (Recall)}} \\
        \cmidrule{4-7}
        & & & \textbf{Category} & \textbf{Quantity} & \textbf{Location} & \textbf{Relationship} \\
        \midrule
        \checkmark & $\times$ & \checkmark & 0.4408 & 0.1452 & - & 0.1378 \\
        \checkmark & $\times$ & $\times$ & 0.1158 & 0.0644 & 0.0308 & 0.0015 \\
        \checkmark & \checkmark & $\times$ & - & - & - & - \\
        \checkmark & \checkmark & \checkmark & \textbf{0.6401} & \textbf{0.3255} & \textbf{0.3227} & \textbf{0.1398} \\
        \bottomrule
    \end{tabular}
    }
\end{table}
Furthermore, when both SSC and OCC are ablated and the JCMSC method was used for image transmission and recovery, performance worsened even further. Without structured semantic encoding, the VQALLM (Visual Question Answering-Large Language Model) model struggles to extract meaningful semantic details from raw image data. In contrast, when both SSC and OCC are retained, recall rates improved significantly across all question types, demonstrating that combining structured semantic encoding with open channel coding is essential for preserving semantic information and enhancing VQA performance.
% The ablation study demonstrates the critical role of both the Structured Semantic Encoder (SSC) and the Open Channel Coder (OCC) in ensuring robust VQA performance. When the OCC encoding method was removed and replaced with the Deep-SC approach, the system's ability to recover semantic information was significantly compromised. As shown in Table~\ref{tab:Deep-SC}, recall rates for "Category," "Quantity," and "Relationship" questions dropped notably. This indicates that without OCC, the semantic information propagation becomes ineffective, hindering the overall system performance.

% When both the SSC and OCC were ablated, the system reverted to using JCMSC for image transmission and recovery. The results showed a marked performance decline, especially in the recall of location and relationship-related questions. This approach lacked structured semantic encoding, making it difficult for the VQALLM model to extract meaningful semantic details from the raw image. In contrast, when both SSC and OCC were used, the system achieved significantly higher recall rates across all question types. This confirms that combining structured semantic encoding with open channel coding enhances the VQA task by preserving deeper semantic information, leading to improved question answering performance.

\subsubsection{Performance Comparison of Different Large Language Models for VQA at the Receiver End} 

In the ablation study, we evaluate the performance of four different large VQALLM models—Llama-3.1, Claude-3.5, GPT-4o, and Qwen-plus—at the receiver end for VQA tasks, with the SNR fixed at 18 dB. The results shown in Table~\ref{tab:LLM} reveal that the choice of VQALLM model causes only minor fluctuations in performance, with no significant impact on overall recall or F1-scores across the attributes of Category, Quantity, Location, and Relationship.

\begin{figure*}[!t]
    \centering
    \includegraphics[width=0.85\linewidth]{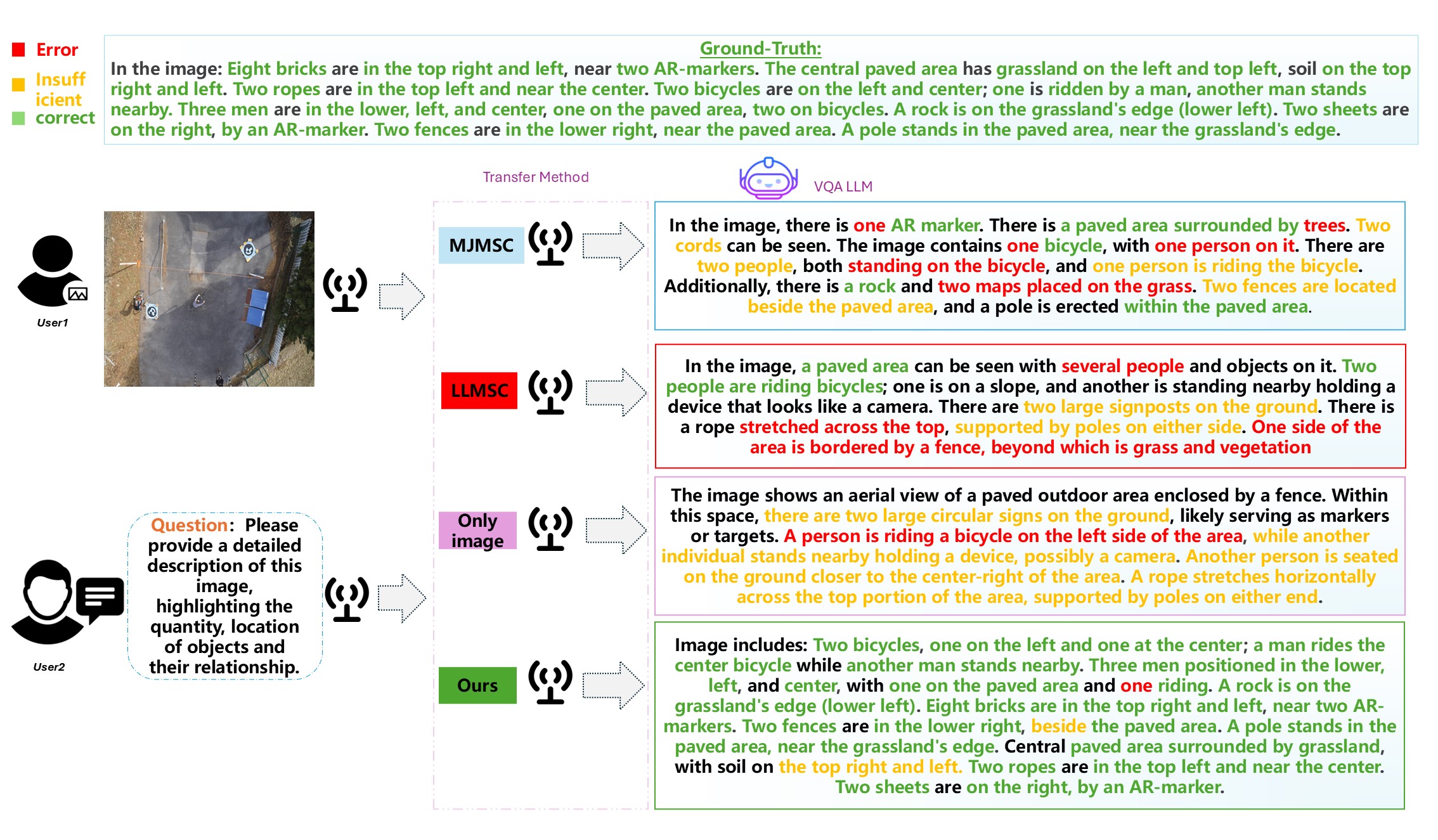}
    \vspace{-0.5cm}
    \caption{Visualization comparing the differences between our method and other methods.}
    % \caption{This is a visualization that clearly demonstrates the differences between our method and other methods.}
    \label{fig:visualization}
\end{figure*}

% \vspace{-0.5cm}
\subsection{Average Number Of Transmitted Symbols}
To assess the efficiency of different image transmission and semantic encoding methods, we compare the average number of transmitted symbols for each approach. A lower symbol count indicates more efficient data transmission, which is vital in bandwidth-limited or efficiency-focused scenarios. This comparison highlights how our method reduces data overhead while maintaining semantic integrity.
\begin{table}[h]
\centering
\caption{Average Number of Transmitted Symbols for Different Methods}
\vspace{-0.25cm}
\label{tab:Transmitted Symbols}
\resizebox{\linewidth}{!}{
\begin{tabular}{l|l|r}
\toprule
\textbf{Methods} & \textbf{\makecell{Average Number of\\Transmitted Symbols}} & \textbf{Ratio (\%)} \\
\midrule
Ours / JPEG-LDPC&    \textbf{4,160} / 16,242,744 & \textbf{$2.6 \times 10^{-4}$}\\
Ours / Huffman+RS &  4,160 / \textbf{1,549} & $2.6856 \times 10^{2}$ \\
Ours / 5bit+RS &   \textbf{4,160} / 4,288 & $9.7015 \times 10^{1}$\\
Ours / MJCMSC &  \textbf{4,160} / 6,472 & $6.4277 \times 10^{1}$\\
Ours / LLMSC &   \textbf{4,160} / 4,220 & $9.8578\times 10^{1}$ \\
\bottomrule
\end{tabular}
}
\end{table}

Table~\ref{tab:Transmitted Symbols} presents a comparison of the average number of transmitted symbols across various methods. All methods utilize 16QAM modulation to ensure fairness, and the JPEG+LPDC method employs an LPDC coding rate of 1/2. The results show that methods involving Scene Graph Generation (SGG) information, such as Huffman+RS, LLM-SC, and our method, transmit significantly fewer symbols than traditional methods like MJCMSC and JPEG+LPDC. Notably, our method requires only 4160 symbols, which is approximately $2.6 \times 10^{-4}$\% of the symbol count needed by JPEG-LDPC, due to its highly compact semantic encoding and the efficient OpenSC method. This demonstrates the superior efficiency of our approach in reducing transmission costs without compromising the quality of the transmitted semantic information.
% Table~\ref{tab:Transmitted Symbols} provides a comparative analysis of the number of transmitted symbols between the proposed method and other methods. It calculates the average number of transmitted symbols corresponding to the textual description (or the image itself) of a single image across the entire dataset.
% From the table, all methods adopt 16QAM modulation to ensure a fair comparison. In the JPEG+LPDC method, the LPDC coding rate is set to 1/2. It can be observed that the methods used for transmitting SGG information, such as Huffman+RS, LLM-SC, and our method, transmit significantly fewer symbols compared to the MJCMSC and the traditional JPEG+LPDC methods for transmitting images.
% Our method has the second fewest transmitted symbols, with only  $2.6 \times 10^{-4}$\% of the symbol count of traditional methods. This is because our method highly semantically encodes the image, making the image's semantics highly integrated. At the same time, the OpenSC encoding method also effectively reduces the number of symbols, which makes our method very close to the symbol count of pure text transmission.

\subsection{Complexity Analysis}
% Table~\ref{tab:complex} compares the computational complexity of LLM-SC and DeepSC based on the average runtime of each item. Our method requires less time compared to MJMSC but is more time-consuming than other methods. This is because the LLM has numerous parameters; however, the introduction of the LLM also eliminates the need for both parties to establish local knowledge bases, requiring only the use of the LLM's public vocabulary list, which reduces the complexity of setting up local knowledge bases. Additionally, with the advancement of computing power, the runtime will further decrease, which will help reduce the computational. complexity of our method.
%%%
The complexity analysis is crucial to assess the computational efficiency of the proposed OpenSC method in comparison to other existing approaches. While our previous experiments highlighted its performance, understanding the computational cost is essential for practical applications.
% \vspace{-0.25cm}
\begin{table}[h]
    \centering
    % \renewcommand{\arraystretch}{1.5}
    % \setlength\tabcolsep{3pt}
    % \caption{Indicating Performance Across Different Large Language Models for Each Attribute}
    \caption{Performance Comparison Across Different Large Language Models for Each Attribute.}
    \vspace{-0.3cm}
    \label{tab:LLM}
    \resizebox{\linewidth}{!}{ % Resizes the table to fit the text width
    \begin{tabular}{ccccccc}
        \toprule
        \multirow{2}{*}{\textbf{VQA LLM}} & \multirow{2}{*}{\textbf{SSC}} & \multirow{2}{*}{\textbf{OCC}} & \multicolumn{2}{c}{\textbf{Category}} & \multicolumn{2}{c}{\textbf{Quantity}}\\
        % \cmidrule{4-11}
        & & & \textbf{Recall} & \textbf{F1 Score}& \textbf{Recall} & \textbf{F1 Score}\\
        \midrule
        llama-3.1~\cite{dubey2024llama} & \checkmark & \checkmark &0.6239& 0.7056 & 0.3089 & 0.3141  \\
        Claude-3.5~\cite{TheC3} & \checkmark & \checkmark  &0.6380& 0.6943& 0.3162 & 0.3036 \\
        GPT-4o~\cite{Achiam2023GPT4TR} & \checkmark & \checkmark & 0.6332& 0.7011 & 0.3367&0.3348  \\
        Qwen-plus~\cite{Bai2023QwenTR} & \checkmark & \checkmark &0.6401& 0.7042 & 0.3255&0.3281 \\\hline
%%%
\multirow{2}{*}{\textbf{VQA LLM}} & \multirow{2}{*}{\textbf{SSC}} & \multirow{2}{*}{\textbf{OCC}} & \multicolumn{2}{c}{\textbf{Location}}& \multicolumn{2}{c}{\textbf{Relationship}}\\
        % \cmidrule{4-11}
        & & & \textbf{Recall} & \textbf{F1 Score}& \textbf{Recall} & \textbf{F1 Score} \\\hline
llama-3.1~\cite{dubey2024llama} & \checkmark & \checkmark &0.3018& 0.2947& 0.1244 & 0.1266 \\
        Claude-3.5~\cite{TheC3} & \checkmark & \checkmark  & 0.3241&0.3373 &0.1187& 0.1210\\
        GPT-4o~\cite{Achiam2023GPT4TR} & \checkmark & \checkmark & 0.3184& 0.3044 & 0.1371&0.1204 \\
        *Qwen-plus~\cite{Bai2023QwenTR} & \checkmark & \checkmark &0.3227& 0.3219 &0.1398& 0.1281  \\
        \bottomrule
    \end{tabular}
    }
    \footnotesize{*Our proposed OpenSC utilizes Qwen-Plus for VQALLM. The main experiments in this paper use Qwen-Plus for VQALLM.}
\end{table}

\begin{table}[htbp]
    \centering
    \caption{Average Processing Time (ms/Item) for Different Methods}
    \vspace{-0.3cm}
    \label{tab:complex}
      \resizebox{\linewidth}{!}{
    \begin{tabular}{c|cccc}
\toprule
        \textbf{Method} & \textbf{DeepSC} & \textbf{MJCMSC}& \textbf{LLMSC}& \textbf{OpenSC (Ours)}  \\
        \hline
        \textbf{Time (ms/Item)} & 35.99 & 1,210.23 & 143.97 & 158.43  \\
\bottomrule
    \end{tabular}
    }
\end{table}

Table~\ref{tab:complex} compares the average processing time per item for OpenSC and other methods. OpenSC requires 158.43 ms, slightly more than LLMSC (143.97 ms) but much faster than MJCMSC (1,210.23 ms). DeepSC is the fastest at 35.99 ms. The higher runtime of OpenSC is mainly due to the large number of parameters in the LLM, though it benefits from eliminating the need for local knowledge base setup. With advancements in computing power, the processing time for OpenSC is expected to further decrease, improving its efficiency.

% \linespread{0.5}
\subsection{Visualization}
The visualization is conducted to highlight the superiority of our method in both VQA tasks and detailed image description generation. As shown in Figure~\ref{fig:visualization}, our method produces a comprehensive image description with only three minor errors, while other methods exhibit more significant errors and semantic loss. This demonstrates the effectiveness of our approach in preserving the full semantic context of the image. Additionally, the integration of RAG~\cite{gao2023retrieval} enhances the precision and detail of our responses, allowing for more accurate and diverse interpretations compared to traditional VQA methods.
% In Figure~\ref{fig:visualization}, we present a visualization that clearly demonstrates the differences between our method and other approaches. The question posed for this visualization is: ``Please provide a detailed description of this image, highlighting the quantity, location of objects, and their relationships." This illustrates that our method is not only effective for VQA tasks but also excels at generating detailed image descriptions.

% As shown, when compared to the Ground-Truth, our method provides a comprehensive image description with only three minor errors, while other methods exhibit more errors and semantic loss. This highlights the robustness of our approach in preserving the full semantic context of the image. Moreover, the integration of RAG~\cite{gao2023retrieval} further improves the accuracy and detail of our answers, enabling more precise responses than traditional VQA methods and allowing for a broader range of query interpretations.
\section{Conclusion}
In conclusion, our propose semantic communication system enhances 6G networks by integrating Large Language Models (LLMs) with structured semantic encoding through scene graphs. This approach improves interpretability, reduces redundancy, and adapts to channel variations in real-time, optimizing transmission efficiency. By focusing on critical objects and relationships, it boosts performance in tasks like Visual Question Answering (VQA). Experimental results validate significant improvements in both semantic understanding and transmission efficiency, advancing multimodal semantic communication and providing a foundation for future adaptive communication systems.

\bibliographystyle{unsrt}
\bibliography{sample-base} %这里的sample-base是你参考

\end{document}